\documentclass[aps,prl,showpacs,twocolumn,superscriptaddress]{revtex4-1}

\usepackage{amsmath}    
\usepackage{graphicx}   
\usepackage{verbatim}   
\usepackage{color}      
\usepackage{subfigure}  
\usepackage{hyperref}   
\usepackage{ulem}
\begin{document}

\title{Microwave control of trapped-ion motion assisted by a running optical lattice}
\author{Shiqian Ding}
\author{Huanqian Loh} 
\author{Roland Hablutzel}
\affiliation{Centre for Quantum Technologies, National University of Singapore, 3 Science Dr 2, 117543, Singapore}
\author{Meng Gao}
\affiliation{Centre for Quantum Technologies, National University of Singapore, 3 Science Dr 2, 117543, Singapore}
\affiliation{Department of Physics, National University of Singapore, 2 Science Dr 3, 117551, Singapore}
\author{Gleb Maslennikov}
\affiliation{Centre for Quantum Technologies, National University of Singapore, 3 Science Dr 2, 117543, Singapore}
\author{Dzmitry Matsukevich}
\affiliation{Centre for Quantum Technologies, National University of Singapore, 3 Science Dr 2, 117543, Singapore}
\affiliation{Department of Physics, National University of Singapore, 2 Science Dr 3, 117551, Singapore}

\date{\today}

\begin{abstract}
We experimentally demonstrate microwave control of the motional state of a trapped 
ion placed in a state-dependent potential generated by a running optical lattice. 
Both the optical lattice depth and the running lattice frequency provide tunability 
of the spin-motion coupling strength. The spin-motional coupling is exploited 
to demonstrate sideband cooling of a $^{171}$Yb$^+$ ion to the ground state of motion. 
\end{abstract}


\maketitle
Control of atomic and molecular motion in the quantum regime is crucial for quantum
information processing, quantum simulation and metrology 
\cite{wineland_blatt_review, wineland_review}. 
The ability to coherently couple the internal state of trapped ions and their motion, 
for example, paved the way
for the demonstration of quantum logic gates \cite{monroe_gate} 
and precision metrology based on quantum logic spectroscopy
\cite{Schmidt_2005, Rosenband_2010}. Traditionally, coupling between internal and motional states 
is achieved by illuminating ions with laser light \cite{wineland_review}. 
The lasers in this case perform two functions. They resonantly couple quantum states 
of the ions, for example, via two-photon stimulated
Raman transitions. They also produce state-dependent potentials that change on 
a scale comparable to the laser wavelength, giving rise to a spin-dependent force. 
As a result, a change of the internal 
state of an ion is accompanied by a change of its motional state. 

Alternative to optical lasers, microwave radiation can be used to drive transitions between 
hyperfine states and shift the ions' energy levels. However, due to 
its long wavelength, the microwave state-dependent potential changes over a significantly larger distance 
and the spin-dependent force is therefore usually weak. The gradient of the 
microwave field can however be significantly enhanced in the near-field regime, 
where microwaves are applied directly to the electrodes of a microfabricated ion trap, leading to spin-motion coupling
\cite{oscillatingproposal,oscillatingdemenstration}. 

The shift of the energy levels and the transition 
between them can also come from two separate 
physical processes, and these combinations already provide benefits for several applications. 
For example, a state-dependent potential can come from the gradient of a static magnetic field, whereas transitions 
can be driven by on-resonant microwave radiation, 
offering  a new approach for quantum information processing with trapped ions \cite{Wunderlich_2001, staticpdemenstration, staticpdemenstration2}. 
Spin-motion coupling was also demonstrated for neutral atoms trapped in a spin-dependent 
optical lattice potential and irradiated by a spatially uniform microwave 
field, leading to a simple scheme for the sideband cooling of atoms to the ground state of motion 
\cite{Meschede_prl_2009,Weiss_2012}. 
An extension of this scheme has also been proposed for performing
quantum logic spectroscopy with molecular ions \cite{microwavemolecule}. However, in the case of trapped ions, interferometric stability between the optical lattice and the ion position is required. Achieving this is a technically challenging task, which usually requires active stabilization of the beam path~\cite{Hume_2011} or 
 integrating the ion trap with an optical cavity \cite{Drewsen_2012,Schaetz_2012,Vuletic_2013,Walther_2001}. 

In this Letter, we experimentally demonstrate microwave coupling between the internal and 
motional states of a trapped $^{171}$Yb$^+$ ion placed in a running spin-dependent optical 
lattice. 
We use this coupling to achieve resolved sideband cooling of a single trapped ion to the ground state 
of motion. Use of the running optical lattice eliminates the requirement for interferometric stability during the entire experiment
(several hours). Instead, the relative phase between two optical beams should be stable for the duration of 
a single experimental cycle only (less than 1 ms), a requirement that is much easier to satisfy in the laboratory. 
In addition, driving internal transitions separated by 12.6~GHz directly with microwaves as opposed to Raman lasers removes the need for high frequency optical modulators.

\begin{figure}[b]
\centering
\includegraphics[width=\columnwidth]{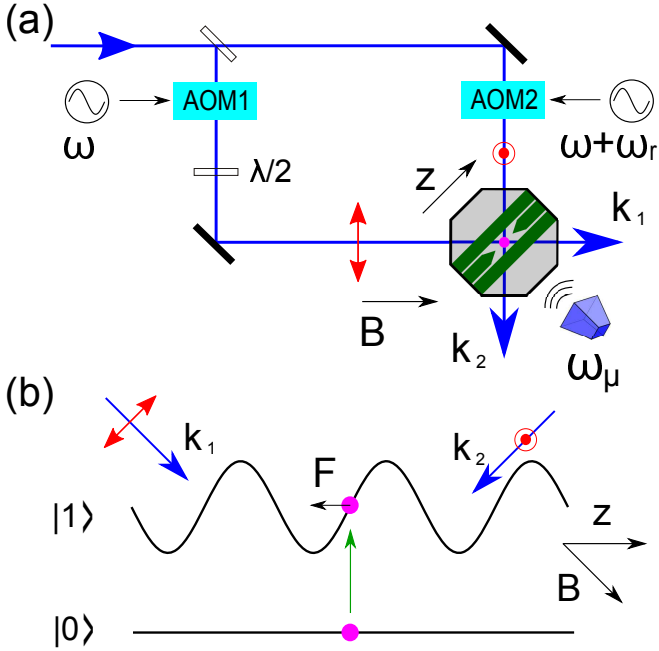}
\caption{\label{fig:setup} (Color online.) (a) Schematic of the experimental setup.  
Two laser beams of wavevectors $\mathbf{k}_1$, $\mathbf{k}_2$ and frequency difference $\omega_r$ are overlapped 
onto the trapped ${^{171}}$Yb$^+$ ion (pink dot) to form an optical lattice 
of running wave frequency $\omega_r$ along $\hat{z}$ direction. 
The red arrow or dot inside a circle represents the polarization of each lattice beam. 
(b) AC Stark shift for the $|1\rangle$ and $|0\rangle$ spin states in the optical lattice at a given instant of time. 
Spin-motion coupling is effected by the optical lattice imparting a kick ($\mathbf{F}$) onto the ion 
upon the microwave field flipping the ion spin state.}
\end{figure} 

A schematic of the experimental setup is shown in Fig. \ref{fig:setup}. 
The single $^{171}$Yb$^+$ ion is
confined in a four-rod linear rf-Paul trap with the secular trapping frequencies  $(\omega_x, 
\omega_y, \omega_z)= 2\pi *(0.91, 0.97, 0.79)$~MHz. 
For Doppler cooling, the ion is illuminated by a 369.53 nm laser of intensity 3~W/cm$^2$ and red-detuned
from the $^{2}S_{1/2},F=1 \rightarrow$ $^{2}P_{1/2},F=0$ transition. 
A 935.19 nm laser of intensity 25~W/cm$^2$ is employed to repump the ion from the 
long-lived  $^{2}D_{3/2}$ state. A magnetic field of 5.5 G is applied at $45\,^{\circ}$ 
angle with respect to the optical lattice axis ($\hat{z}$) to destabilize the dark states in the $^{2}S_{1/2}$ manifold. The 
standard optical pumping and resonance fluorescence state detection techniques as described in \cite{2007PRA} are used in the experiment 
to initialize and detect the ion's quantum state. 

To produce the optical lattice beams, the output of a 1.6~W mode-locked Ti:Sapphire laser (pulse 
duration 3~ps, repetition rate 76~MHz) is frequency-doubled by 
a Lithium triborate crystal, generating about 200~mW of 
377.2~nm light. 
The resulting beam is then split into two, sent through two separate acousto-optical modulators (AOMs), 
and focused to a beam waist of 15~$\mu$m at the ion position from two 
orthogonal directions that form 45$^\circ$ and 135$^\circ$ angles respectively with the $\hat{z}$-axis.
The path lengths of the two beams are matched to a precision much better 
than the picosecond pulse length. The beams  
interfere at the ion position to generate an optical lattice superimposed on the harmonic ion trap potential. 
The polarizations of the two beams are linear and mutually orthogonal. 
In this configuration, the polarization in the optical 
lattice changes from linear to circular to orthogonal 
linear to opposite circular and back to linear within $\sqrt{2}\lambda/2 = 2 \pi / \Delta k$, 
giving rise to a strong spatial 
dependence of the differential Stark shift between the $|0\rangle=|S_{1/2}, F=0, m_{F}=0\rangle$ and 
$|1\rangle=|S_{1/2}, F=1, m_{F}=-1\rangle$ states of the trapped ion. Given the repetition rate of the laser is much faster than the trap frequency, only the average optical potential affects the ion motion. Hence, if the driving frequencies of the two AOMs differ by $\omega_r$, we will get effectively the same running optical lattice propagating along the $\hat{z}$ direction as if the pulsed laser is replaced by a continuous wave laser. Care should be taken to choose the repetition rate of the pulsed laser such that the generated frequency comb does not drive any stimulated Raman transitions between hyperfine states of the Yb$^+$ ion \cite{Hayes_2010}.

The microwave radiation at 12.6~GHz, emitted by a microwave horn placed 5~cm away from the trap, is used to drive the magnetic dipole transition between $|0\rangle$ and $|1\rangle$ with a Rabi frequency of $\Omega /2 \pi = 43$~kHz. By measuring the shifts of the 
microwave resonance in the presence of a slowly running optical lattice with $\omega_r \ll \Omega$, we estimate the 
differential ac Stark shift to be $\Delta \omega_0 /2\pi= \pm 310(10)$~kHz.

The interaction of a trapped ion with the running optical lattice and microwave 
field of frequency $\omega_\mu$ 
can be described by a Hamiltonian of the form $H = H_0 + V(t)$, where 
\begin{eqnarray}
H_0 &=& \hbar \omega_z a^{\dagger} a  + \frac{1}{2} \hbar \omega_0 \sigma_z,  \\
V(t) &=& \nonumber
\frac{1}{2} \hbar  \sigma_z \delta \omega_0 (z, t) 
 + \frac{1}{2} \hbar \Omega (\sigma_{+} + \sigma_{-}) 
 (e^{-i \omega_\mu t} +  e^{i \omega_\mu t} ). \nonumber
\end{eqnarray}
Here $\hbar \omega_0$ is the energy difference between the ion internal states $|0\rangle$ and $|
1\rangle$, 
$\sigma_z = {|1\rangle\langle 1| - |0\rangle\langle 0|}$, 
$\sigma_+ = {|1\rangle\langle 0|}$
$\sigma_- = {|0\rangle\langle 1|}$, 
$a^\dagger$ and $a$ are creation and annihilation operators for the ion motional mode, 
$\omega_z$ is the secular trap frequency,
$\Omega$ is the microwave transition Rabi frequency,
and $\delta \omega_0 (z, t) = \Delta \omega_0 \sin(\Delta k z - \omega_r t) $ is the differential Stark shift of 
the ion energy levels in the presence of the running optical lattice. In the Lamb-Dicke regime, near the ion equilibrium position $z=0$, 
we can keep only terms to the first order in $z$ in an expansion of the differential Stark shift expression, i.e.
\begin{equation}
\delta \omega_0 (z, t) = \Delta \omega_0 (-\sin \omega_r t + \eta (a^{\dagger} + a) \cos 
\omega_r t). 
\end{equation}
Here $\eta = \Delta k z_0$ is the Lamb-Dicke parameter, $z_0 = \sqrt{\hbar / 2 m \omega_z}$ is the 
spread of the ion wave function along the $\hat{z}$ direction, 
and $m$ is the ion mass.

After transforming the Hamiltonian to the interaction picture 
$H_i(t) = e^{i H_0 t / \hbar} V(t) e^{-i H_0 t / \hbar} $ 
and using the rotating wave approximation for the microwave interaction, we obtain
\begin{eqnarray}
H_i(t) &=& \frac{1}{4} \hbar \Delta \omega_0 \eta \sigma_z a 
  ( e^{-i (\omega_z + \omega_r) t } +  e^{-i (\omega_z - \omega_r) t }) \\
  && - \frac{i}{4} \hbar \Delta \omega_0  \sigma_z e^{-i \omega_r t} 
  + \frac{1}{2} \hbar \Omega \sigma_{+} e^{-i \delta t} + H.c. \, , \nonumber 
\end{eqnarray}
where $\delta = \omega_\mu - \omega_0$ is the detuning of the microwave field from the 
$|0\rangle \rightarrow |1\rangle$ transition. 

Following Ref. \cite{effective_h} and dropping fast oscillating terms, 
the effective Hamiltonian for the detuning $\delta = \pm \omega_r$ is given by 
\begin{equation}
H_{st} =  i \hbar \Delta \omega_0 \Omega (\sigma_{+} - \sigma_{-}) / (4 \omega_r),
\end{equation}
which corresponds to a change in the ion internal spin state only. In a similar way, the effective Hamiltonian for the sidebands at detuning $\delta = -(\omega_z \pm \omega_r)$ can be written as 
\begin{equation}
H_{rsb} = - \frac{\hbar \Delta \omega_0 \eta \Omega}{4 (\omega_z \pm \omega_r) } 
\left( a \sigma_{+} + a^{\dagger} \sigma_{-} \right).
\label{eq:rsb} 
\end{equation} 
When tuned to this red sideband \cite{wineland_bible}, the microwave field drives the $|0\rangle \rightarrow |1\rangle$ transition while destroying a phonon in the ion motional mode.
When the condition $\delta = \omega_z \pm \omega_r$ is 
satisfied, the effective Hamiltonian given by
\begin{equation}
H_{bsb} = \frac{\hbar \Delta \omega_0 \eta \Omega}{4 (\omega_z \pm \omega_r) } 
\left( a \sigma_{-} + a^{\dagger} \sigma_{+} \right)
\label{eq:bsb}
\end{equation}
describes blue sidebands, i.e.\ the $|0\rangle \rightarrow |1\rangle$ transition 
is accompanied by the creation of a phonon in the ion motional mode. 
The spin-motion couplings induced by these 
terms are analogous to the more familiar red and blue sidebands driven by a pair of 
Raman lasers \cite{wineland_bible}, but with the 
Lamb-Dicke parameter replaced by $\eta^{eff}_{\pm} = \eta \Delta \omega_0 / (2 (\omega_z \pm \omega_r) )$.
\begin{figure}[tb]
\centering
\includegraphics[width=\columnwidth]{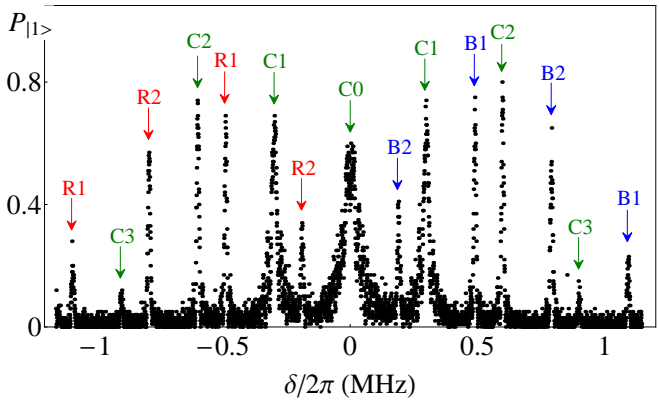}
\caption{\label{fig:sb} (Color online.)  
Probability of the $|0\rangle \rightarrow |1\rangle$ transition after a 75~$\mu$s microwave pulse as a function of microwave detuning $\delta$. Every point is the average of 100 measurements. The peak labeled C0 corresponds to the carrier microwave transition, while R1 and B1 correspond to the red and blue motional sidebands respectively. Peaks labeled Cn, Rn, Bn contain contributions from higher order terms that scale as $(\Delta \omega_0 / \omega_z)^n$.  }
\end{figure}

The measured probability of the $|0\rangle \rightarrow |1\rangle$ transition as a function of 
the microwave detuning $\delta$ is shown in Fig.~\ref{fig:sb} for $\omega_r / 2 \pi= 300$~kHz. 
Three kinds of transitions can be identified in this graph. 
The main carrier peak (C0) corresponds to the transition between the internal states 
of the ion without changing its motional state. It
is accompanied by a series of smaller peaks C$n$ separated by $\omega_r$ that correspond
to higher order terms that scale as $(\Delta \omega_0 / \omega_r)^n$. These peaks are due to periodic changes of the 
resonance frequency by the Stark shift and are mathematically similar to
the sidebands generated by the ion micromotion \cite{wineland_review}. The B1 (R1) sidebands at the 
detunings $\omega_z \pm \omega_r$ and $-(\omega_z \pm \omega_r)$ respectively 
correspond to transitions that change both internal 
and motional states of the ion, and are described by the effective Hamiltonian $H_{bsb}$ 
($H_{rsb}$). As in the case of the carrier, these peaks are accompanied by sidebands of higher order in  $\Delta \omega_0 / \omega_r$. 

\begin{figure}[tb]
\centering
\includegraphics[width=\columnwidth]{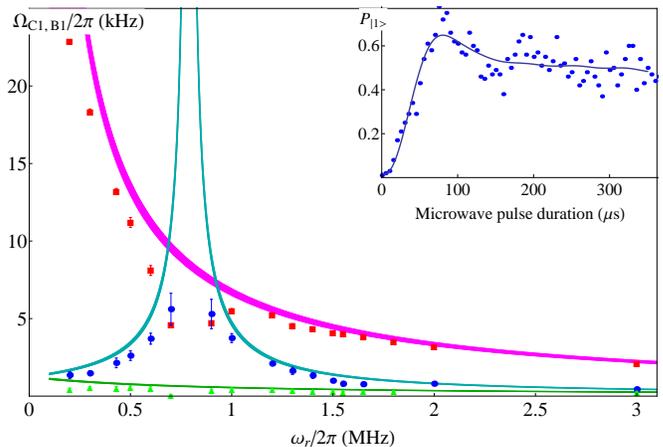}
\caption{\label{fig:bsbrabiscan}
(Color online.) Rabi frequencies for the microwave detuning  $\delta = \omega_r$ (red squares, C1), $\delta = \omega_z - \omega_r$ (blue dots, B1) and $\delta = \omega_z + \omega_r$ (green triangles, B1), plotted as a function of the running lattice frequency $\omega_r$. The solid lines are theoretical predictions with no free parameters, with their thickness representing prediction uncertainty. The Rabi frequencies $\Omega_{B1}$ are obtained from the internal state evolution of the ion as shown in the inset for $\delta = \omega_z - \omega_r=2\pi*490$ kHz. The Rabi oscillation data is fit assuming a thermal distribution of phonons after Doppler cooling.}
\end{figure}

For a given running lattice frequency $\omega_r$ and microwave detuning $\delta = \omega_z \pm \omega_r$, the internal-state evolution of the Doppler-cooled ion is measured (see Fig.~\ref{fig:bsbrabiscan} inset). To extract the Rabi frequency $\eta^{eff}_{\pm} \Omega$ , it is then fit to a weighted average of Rabi oscillations assuming a thermal distribution of phonons \cite{1996PRLNonclassicalstate}. 
The ion temperature corresponds to an average phonon number $\langle n \rangle = 18(2)$. 
As plotted in Fig.~\ref{fig:bsbrabiscan} (blue dots and green triangles), the Rabi frequencies for the two B1 sidebands are generally asymmetric over a wide range of $\omega_r$. The blue and green solid lines are theory curves, described by 
$\Omega_{B1} = \eta^{eff}_{\pm} \Omega $
with no free parameters. 
Also depicted in Fig.~\ref{fig:bsbrabiscan} are the experimental (red squares) and theoretical (red line) Rabi frequencies 
for the C1 sideband at $\delta = \omega_r$, where the former one is extracted by a fit to an exponentially decaying sinusoidal function, while the latter one is given by $\Omega_{C1} = \Delta \omega_0 \Omega / 2 \omega_r$. The theoretical prediction uncertainty, which comes from uncertainties in $\Omega$, $\eta$ and $\Delta \omega_0$, is represented by solid lines' thickness in Fig.~\ref{fig:bsbrabiscan}. As expected from Eq.~(\ref{eq:bsb}), the Rabi frequency of the $\omega_z - \omega_r$ sideband exhibits resonance behavior when $\omega_r$ approaches $\omega_z$. Some quantitative discrepancy between the theoretical prediction and experimental values near the resonance are due to the limitations of our theoretical treatment that does not take into account higher order terms in  $\Delta \omega_0 / \omega_r$. The strong dependence of the R1, B1 sideband Rabi frequencies on $\omega_r$ opens up the possibility of speeding up operations that depend on spin-motion coupling, such as sideband cooling.

\begin{figure}[tb]
\centering
\includegraphics[width=\columnwidth]{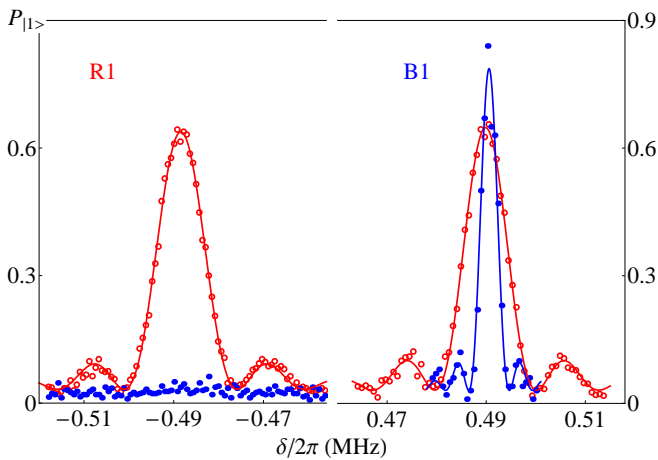}
\caption{\label{fig:sbsbc} (Color online.) Probability of the $|0\rangle \rightarrow |1\rangle$ transition for red sideband (left, R1) and blue sideband (right, B1) at detunings $\delta = \mp(\omega_z - \omega_r)$ before (red open-circles, microwave pulse duration 80 $\mu$s) and after (blue dots, microwave pulse duration 230 $\mu$s) sideband cooling. Suppression of the red sideband after sideband cooling indicates cooling to near the motional ground state.}
\end{figure}
To demonstrate sideband cooling, we revert to the case of $\omega_r/2\pi = 300$~kHz and use the red sideband at the detuning $\delta = \omega_r - \omega_z = -2\pi*490$~kHz (R1). After 1~ms of Doppler cooling, 200 microwave 
pulses at the detuning $\delta$ are applied to the ion. Each pulse is followed by 5~$\mu$s of optical pumping to reinitialize the ion back in the $|0\rangle$ state. 
The duration of the microwave pulse is increased from 60 to 230~$\mu$s in steps of approximately 1 $\mu$s throughout 
the sideband cooling sequence to account for the increased $\pi$ time of the 
sideband transition as the mean phonon number decreases. The corresponding red and blue 
sidebands before and after sideband cooling are shown in Fig. 4. The red sideband after 
sideband cooling is mostly diminished. 
The height of the sideband-cooled blue sideband is limited by decoherence of the $|0\rangle \rightarrow |1\rangle$ transition (coherence time 0.47(4) ms), which is mostly attributed to magnetic field noise.
From the asymmetry of the red and blue sidebands, 
we estimate the average number of phonons in the motional mode of the ion to be $\langle n \rangle = 0.02\pm 0.04$. In addition, the red sideband at detuning $\delta = - \omega_z = -2\pi*790$~kHz (R2) was used to achieve sideband cooling to $\langle n \rangle = 0.02\pm 0.02$. 

In summary, we realize spin-motion coupling for a single trapped ion using 
a uniform microwave field assisted by a running optical lattice. 
This technically simple scheme allows us to cool the ion to the ground state of 
motion. 
The running lattice-microwave combination offers two tiers of flexibility on the spin-motion coupling strength: 
one can either vary the optical lattice depth or tune the running lattice frequency relative to the secular trap frequency. 
The increased flexibility allows for more tunable quantum logic gates and spin-spin couplings mediated by phonon modes, 
where the latter can be exploited in quantum simulation 
\cite{staticpdemenstration2,Kim_2009,blatt_simulations_2012} or studies of transport properties of the ions interacting with the optical lattice
\cite{Drewsen_2012,Schaetz_2012,Vuletic_2013,Haffner_2011,Plenio_2013}. This technique also paves the way for molecular ion quantum logic spectroscopy \cite{ourpaper,didi'spaper}, by which the rich internal structure of molecular ions and their motion can be accessed \cite{microwavemolecule}.

This research was supported by the National Research
Foundation and the Ministry of Education of Singapore.

%


\end{document}